\documentclass{PoS}
 \def\cB{{\cal B}}

\newcommand{\nn}{\nonumber}
\newcommand{\be}{\begin{equation}}
\newcommand{\ee}{\end{equation}}
\newcommand{\bea}{\begin{eqnarray}}
\newcommand{\eea}{\end{eqnarray}}
\newcommand{\RK} {R^{\mu/e}_K}
\newcommand{\RKs} {R^{\mu/e}_{K^\ast}}
\newcommand{\RKss} {R^{\mu/e}_{K^{(\ast)}}}

\newcommand{ \mysmall}[1]{\scriptscriptstyle #1}

\title{B-anomalies related to leptons and lepton flavour violation: new directions in model building}

\ShortTitle{B-anomalies and LFV}

\author{\speaker{Ferruccio Feruglio}\\
        Dipartimento di Fisica e Astronomia `G.~Galilei', Universit\`a di Padova\\
INFN, Sezione di Padova, Via Marzolo~8, I-35131 Padua, Italy\\
        E-mail: \email{feruglio@pd.infn.it}}

\abstract{$B$-decays mediated by both charged currents and neutral currents 
have provided hints of violation of lepton flavour universality. 
By assuming an explanation of these anomalies in terms of new physics at 
the TeV scale affecting the third fermion generation, we review the relevant constraints.
In particular we illustrate the effects of electroweak radiative 
corrections and their impact on a consistent interpretation of the data.
The simplest scenario is ruled out and we briefly discuss the modifications required
to simultaneously describe the whole set of anomalies.}

\FullConference{The International Conference on B-Physics at Frontier Machines - BEAUTY2018\\
		6-11 May, 2018\\
		La Biodola, Elba Island, Italy}

\begin{document}
\section{Introduction}
$B$-meson decays have access to charged leptons of the three generations.
This fact, together with the distinctive experimental signatures of the $B$-mesons,
makes $B$-semileptonic transitions a good playground to test lepton flavour universality (LFU).
In the last few years experiments have accumulated hints of violation of LFU, both in neutral currents (NC) 
transitions and in charged current (CC) \cite{Simonetto} ones. 
Particularly relevant are the NC ratios
\bea
\RKs &=&  \left. \frac{ \cB(B \to K^* \mu \bar{\mu})_{\rm exp} }{ \cB(B \to K^* e \bar{e} )_{\rm exp} } \right|_{q^2\in[1.1,6]{\rm GeV}} =  0. 685 {}^{+0.113}_{-0.069} \pm 0.047~,
\label{eq:RKSexp} \\
\RK &=&  \left. \frac{ \cB(B \to K \mu \bar{\mu})_{\rm exp} }{ \cB(B \to K e \bar{e} )_{\rm exp} } \right|_{q^2\in[1,6]{\rm GeV}} =  0. 745 {}^{+0.090}_{-0.074} \pm 0.036~,
\label{eq:RKexp}
\eea
based on combination of LHCb data~\cite{Aaij:2017vbb} with the SM expectation $\RKss =1.00 \pm 0.01$~\cite{Bordone:2016gaq,Hiller:2003js}, and the CC ratios \cite{Owen}
\bea
R^{\tau/\ell}_{D^*} &=&\frac{ \cB(B \to D^* \tau \overline{\nu})_{\rm exp}/\cB(B \to D^* \tau \overline{\nu})_{\rm SM} }{ \cB(B \to D^* \ell \overline{\nu} )_{\rm exp}/ \cB(B \to D^* \ell \overline{\nu} )_{\rm SM} } = 1.23 \pm 0.07~, \label{eq:RDexp}  \\
R^{\tau/\ell}_{D } &=& \frac{ \cB(B \to D  \tau \overline{\nu})_{\rm exp}/\cB(B \to D  \tau \overline{\nu})_{\rm SM} }{ \cB(B \to D  \ell \overline{\nu} )_{\rm exp}/ \cB(B \to D  \ell \overline{\nu} )_{\rm SM} } =   1.34 \pm 0.17~, \label{eq:RDSexp}
\eea
where $\ell=e, \mu$, following from the HFAG averages of Babar, Belle, 
and LHCb data \cite{Owen}, combined with the SM predictions~\cite{Sumensari}.
Beyond the above ratios, where theoretical uncertainties cancel to a large extent, other observables
such as angular distributions and differential rates in NC semileptonic $B$-decays \cite{Wang,Turchikhin,Smith,Puig} can provide additional 
useful information, though their theoretical interpretation is affected by a considerable ambiguity \cite{Ciuchini}.

We recall that in the Standard Model (SM), where we can safely set neutrino masses to zero and the 
lepton mixing matrix to unity to the purpose of this discussion, the following rules apply:
\begin{itemize}
\item[1.] Absence of lepton flavour violation (LFV) in charged lepton transitions.
\item[2.] Violation of LFU fully controlled by the charged lepton masses.
\end{itemize}
The first one is very well verified: no exception is known. The tight bounds that have been set \cite{Glenzinski}, for example,
on the branching ratios of the decays $\mu\to e~\gamma$ and $\mu\to 3 e$ can be converted into strong lower limits
on the scale $\Lambda$ of new physics (NP) that might be implied in these transitions. From $\cB(\mu \to e \gamma)<4.2\times 10^{-13}$ \cite{TheMEG:2016wtm} we get $\Lambda>10^5$ TeV, out of the range directly explorable by the present facilities.
The second rule is also well verified, at the per mil level, in a wide energy range. At the GeV scale it has been tested in many leptonic and semileptonic light-pseudoscalar decays. In leptonic tau decays we have
$R^{\tau/e}_\tau = 1.0060 \pm 0.0030$ and $R^{\tau/\mu}_\tau = 1.0022 \pm 0.0030$~\cite{Pich:2013lsa}, where 
\be
\label{Rtau}
R^{\tau/e}_\tau = \frac{\mathcal{B}(\tau \!\to\! \mu \nu\bar\nu)_{\rm exp}/\mathcal{B}(\tau \!\to\! \mu \nu\bar\nu)_{\rm SM}}{\mathcal{B}(\mu \!\to\! e \nu\bar\nu)_{\rm exp}/\mathcal{B}(\mu \!\to\! e \nu\bar\nu)_{\rm SM}} \qquad
R^{\tau/\mu}_\tau = \frac{\mathcal{B}(\tau \!\to\! e \nu\bar\nu)_{\rm exp}/\mathcal{B}(\tau \!\to\! e \nu\bar\nu)_{\rm SM}}{\mathcal{B}(\mu \!\to\! e \nu\bar\nu)_{\rm exp}/\mathcal{B}(\mu \!\to\! e \nu\bar\nu)_{\rm SM}} \,.
\ee
At the electroweak scale LFU has been tested in leptonic $Z$ decays at LEP~\cite{Agashe:2014kda}:
\be
\label{vecax}
\frac{v_\tau}{v_e} = 0.959\; (29) ~~~~~~~~~~~~~~~~~ \frac{a_\tau}{a_e} = 1.0019\; (15) \, ,
\label{eq:zpole_pdg}
\ee
where $v_f$ and $a_f$ $(f=e,\mu,\tau)$ are the vector and axial-vector couplings of the lepton $f$ to the $Z$ boson.
A long-standing exception to rule 2. is the observed deviation of the anomalous magnetic moment of the muon  \cite{Bennett:2006fi}
from the SM prediction, which can be interpreted as violation of LFU in the electron-muon sector.  This result is waiting  an independent confirmation by the Fermilab Muon $(g-2)$ experiment.

Any violation of 1. and/or 2. entails physics beyond the SM. In most SM extensions, LFV and LFU violation
are strictly related. They often come together, though this is not an unavoidable feature. Coming back to the hints of LFU violation
in $B$ decays, a first important requirement of possible NP scenarios called for their explanation is the compatibility with the existing tests of LFV and LFU. It is difficult to investigate this aspect in full generality. Here we will first focus on a specific scenario,
providing a concrete benchmark. In the concluding remarks we will discuss possible departures from the benchmark.
\section{A Benchmark Scenario}
The framework considered here \cite{Feruglio:2016gvd,Feruglio:2017rjo} is based on two assumptions and one empirical ingredient. First, we assume that NP occurs above the electroweak scale.
This allows us to describe NP effects through a combination
of gauge-invariant dimension-6 semileptonic operators. The requirement of invariance under the SM gauge
group greatly reduces the number of independent parameters at the high scale $\Lambda$.
Second, we assume that NP only affects the third generation, both in the quark and in the lepton sectors.
In this setup the couplings to lighter generations ($c$-quark, $\mu$) are generated by a misalignement between
mass and interactions basis. Out of the eight independent semileptonic operators contributing to NC/CC $B$-decays,
we will first focus on those depending only on left-handed quarks and leptons. This choice is supported by global fits
to NC data \cite{Nazila,Strumia}, though not in a conclusive way. Our benchmark scenario is defined by the effective NP Lagrangian:
\be
{\cal L}_{NP}^0(\Lambda)=\frac{1}{\Lambda^2}
\left(
C_1~ \bar q'_{3L}\gamma^\mu q'_{3L}~ \bar \ell'_{3L}\gamma_\mu \ell'_{3L}+C_3~\bar q'_{3L}\gamma^\mu \tau^a q'_{3L}~ 
\bar \ell'_{3L}\gamma_\mu \tau^a \ell'_{3L}
\right)\,.
\ee
We can move to the mass basis by means of the unitary transformations
\bea
u'_L=V_u u_L~~,~~~~~~~~~~
& 
d'_L=V_d d_L ~~,~~~~~~~~~~
&
V_u^\dagger V_d=V_{CKM}~~,\\
\nu'_L=U_e \nu_L~~,~~~~~~~~~~
&
e'_L=U_e e_L~~,~~~~~~~~~~
&
\eea
where $V_{CKM}$ is the CKM mixing matrix. We get
\bea\label{LNP}
{\cal L}_{NP}^0(\Lambda)&=&\frac{\lambda^e_{kl}}{\Lambda^2}\left[\right.(C_1+C_3)~\lambda^u_{ij}~\bar u_{Li}\gamma^\mu u_{Lj}~ \bar\nu_{Lk}\gamma_\mu \nu_{Ll}+(C_1-C_3)~\lambda^u_{ij}~\bar u_{Li} \gamma^\mu u_{Lj}~\bar e_{Lk}\gamma_\mu e_{Ll}~+\label{lag1}\nn\\
&&(C_1-C_3)~\lambda^d_{ij}~\bar d_{Li} \gamma^\mu d_{Lj}~ \bar\nu_{Lk}\gamma_\mu \nu_{Ll}+(C_1+C_3)~\lambda^d_{ij}~\bar d_{Li} \gamma^\mu d_{Lj}~\bar e_{Lk}\gamma_\mu e_{Ll}~+\\
&&2 C_3~\left(\lambda^{ud}_{ij}~\bar u_{Li} \gamma^\mu d_{Lj}~\bar e_{Lk}\gamma_\mu \nu_{Ll}+h.c.\right)\left.\right]~,\nn
\eea
where the $\lambda$ matrices reflect the composition of the third generation in terms of mass eigenstates:
\be
\lambda^u_{ij}=V_{u3i}^*V_{u3j} ~~~~~~~~~~~~			
\lambda^d_{ij}=V_{d3i}^*V_{d3j} ~~~~~~~~~~~~	
\lambda^{ud}_{ij}=V_{u3i}^*V_{d3j} ~~~~~~~~~~~~	
\lambda^e_{ij}=U_{e3i}^*U_{e3j} ~~~.	
\ee
Not all these matrices are independent, since the following relations hold:
$\lambda^u=V_{CKM} \lambda^d V_{CKM}^\dagger$ and $\lambda^{ud}=V_{CKM}\lambda^d$. 
To minimally parametrize $\lambda^e$ and $\lambda^d$ we assume that $e_{3L}$ and $d_{3L}$ have vanishing components along the lightest state: $\lambda^f_{22}=\sin^2\theta_f$, $\lambda^f_{33}=\cos^2\theta_f$
and $\lambda^f_{23}=\lambda^f_{32}=\sin\theta_f\cos\theta_f$ $(f=e,d)$, the remaining elements being zero.
Thus the benchmark scenario depends on a minimal set of four parameters, the Wilson coefficients $C_{1,3}/\Lambda^2$,
and the mixing angles $\theta_{e,d}$. Both NC and CC anomalies can be explained provided $C_{1,3}/\Lambda^2$ are of order one TeV$^{-2}$ and $\theta_e\approx 0.3$ and $\theta_d\approx 0.01$ \cite{Calibbi:2015kma}. To remain in a perturbative regime
the scale $\Lambda$ cannot be larger than approximately 1 TeV.
These values efficiently reproduce the relative suppression between NC and CC transitions
which, in the SM, arise at one-loop level and at the tree level, respectively. Indeed, in the small angle approximation, the deviations $\RK-1$ and $\RKs-1$
scale as $(C_1+C_3)\theta_d\theta_e^2$, while $R^{\tau/\ell}_{D^{(*)}}-1$ is proportional to $C_3$. The same parametric behavior is expected in the corresponding purely leptonic NC and CC $B$-transitions. For instance, the branching ratio of $B_s\to \mu^+\mu^-$ \cite{Fedi,Rama}
should deviate from the SM predictions by an amount proportional to $(C_1+C_3)\theta_d\theta_e^2$, expected to be of 
relative order 0.1. Similarly, the deviation from one of the ratio 
\be
R^{\tau/\mu}_{B\tau\nu} = 
\frac{\mathcal{B}(B \to \tau\nu)_{\rm exp}/\mathcal{B}(B \to \tau\nu)_{\rm SM}}{\mathcal{B}(B \to \mu\nu)_{\rm exp}/\mathcal{B}(B \to \mu\nu)_{\rm SM}} 
\,,
\ee
is proportional to $C_3$ and estimated to be of order 0.1. 
As already observed in~\cite{Calibbi:2015kma}, the process $B\to K^{(*)} \nu\bar\nu$ \cite{Merola,Sandilya} sets relevant constraints on our model.
Defining $R^{\nu\nu}_{K^{(*)}}=\mathcal{B}(B\to K^{(*)} \nu\bar\nu)/\mathcal{B}(B\to K^{(*)} \nu\bar\nu)_{\mysmall\rm SM}$
we find that the deviation from one, in a linearized approximation, is proportional to $(C_1-C_3)\theta_d$ and can be
of order 1. The present upper bounds $R^{\nu\nu}_{K} <  4.3$, $R^{\nu\nu}_{K^*} <  4.4$
provide a significant limitation on the available parameter space, favouring the region where $C_1$ and $C_3$ have the same sign and size.
In the benchmark scenario LFV is strictly related to LFU violation \cite{Glashow:2014iga}. Semileptonic or purely leptonic $B$-decays with $\mu^\pm \tau^\mp$ in the final state are directly related to the anomalous NC transitions. The branching ratio $\mathcal{B}(B\to K \mu^\pm \tau^\mp)$ is proportional to $|(C_1+C_3)\theta_d\theta_e|^2$ and expected to be of order $10^{-(6\div7)}$, below the present experimental bound
$\mathcal{B}(B\to K \mu^\pm \tau^\mp)<4.8\times 10^{-5}$. Moreover the following approximate relations hold:
$\mathcal{B}(B\to K^* \mu^\pm \tau^\mp)\approx 2 \mathcal{B}(B\to K \mu^\pm \tau^\mp)$, $\mathcal{B}(B\to \mu^\pm \tau^\mp)\approx \mathcal{B}(B\to K \mu^\pm \tau^\mp)$. The effective operators of the NP Lagrangian, eq. (\ref{LNP}), are also responsible for an 
overproduction of high-$p_T$ $\tau^+\tau^-$ at hadron colliders \cite{Admir}. The signals depend on the type of mediator whose exchange give rise 
to the effective Lagrangian. For colorless mediators, like for example a $Z'$, the production of $\tau$ pairs proceeds via the $s$-channel
and is sensitive to the $Z'$ width. In the case of leptoquark (LQ) mediators, the production channels are $u$/$t$ and 
the signal is scarcely dependent on the mediator width. In both cases the process is of primary importance to correctly determine
the available parameter space, especially in the context of UV complete models. 
\section{Constraints from Quantum Effects}
Another set of constraints originates from quantum effects. These arise when moving from the high scale $\Lambda$, where a description based on the NP effective Lagrangian ${\cal L}_{NP}^0(\Lambda)$ holds, to the lower scale $\mu\approx 1$ GeV:
\be
{\cal L}_{NP}(\mu)={\cal L}_{NP}^0(\Lambda)+{\tt quantum~corrections}
\ee
Why such corrections, typically of order $\alpha/4\pi\approx 10^{-3}$, should be relevant to our discussion, given the comparatively large size
of the anomalies in $B$-decays? Through the well-known phenomenon of operator mixing, quantum corrections can generate an entirely
new set of operators, absent in ${\cal L}_{NP}^0(\Lambda)$, potentially affecting physical processes other than those discussed above.
Moreover the expected order of magnitude is similar to the accuracy in electroweak precision tests and in other tests of LFU, as we
saw in the first Section. In this respect, also in view of the enhancement proportional to $\log(\Lambda/m_t)$, the effects generated by quantum corrections can reveal crucial to establish the viability of the benchmark model.
In the present context, NP described by semileptonic current$\times$current operators, quantum effects are dominated by
electroweak corrections and standard renormalization group techniques can be applied
\cite{Jenkins:2013wua,Alonso:2013hga}. Starting from the scale $\Lambda$
we run our Lagrangian down to lower energies, until we reach the electroweak scale $m_Z$. Here the new effective
Lagrangian ${\cal L}_{NP}(m_Z)$ includes additional operators responsible for non-universal modification of the $W/Z$ coupling to leptons  \cite{Feruglio:2016gvd,Feruglio:2017rjo}. This can be qualitatively understood by 
closing in a loop the quark lines of the original semileptonic operators 
and attaching a gauge vector boson to them. The important feature of the leading correction is its proportionality to $m_t^2/\Lambda^2$, arising from the top quark circulating in the loop. In terms of the LFU tests related to 
vector and axial-vector $Z$-couplings to leptons we find (choosing hereafter $\Lambda=1$ TeV in $\log(\Lambda/m_t)$):
\bea
\frac{v_\tau}{v_e} & \approx& 1 - 0.05 \, \frac{\left( C_1 - 0.8 \,C_3 \right)}{\Lambda^2({\rm TeV^2})}\,, 
\\
\frac{a_\tau}{a_e} & \approx& 1 - 0.004 \, \frac{\left( C_1 - 0.8 \,C_3 \right)}{\Lambda^2({\rm TeV^2})}\,,
\label{eq:zpole_num}
\eea
showing that the experimental results of eq. (\ref{vecax}) can significantly affect the region of parameter space
relevant to the explanation of the $B$-anomalies. We also find:
\begin{equation}
N_\nu \approx 3 + 0.008 \, \frac{\left( C_1 + 0.8 \,C_3 \right)}{\Lambda^2({\rm TeV^2})}\,,
\end{equation}
to be compared with the experimental result~\cite{Agashe:2014kda} $N_\nu = 2.9840 \pm 0.0082$.
LFV violating decays of the $Z$ boson are expected. We estimate $\mathcal{B}(Z\to \mu^\pm \tau^\mp)\approx 10^{-7}$, still too
far from the LEP bound $\mathcal{B}(Z\to \mu^\pm \tau^\mp)\approx 1.2\times 10^{-5}$. Deviations in $W$ decays are also
predicted, though smaller than the present experimental accuracy.

\begin{figure}[h!]
\centering
\begin{minipage}{0.45\textwidth}
\includegraphics[width=\textwidth]{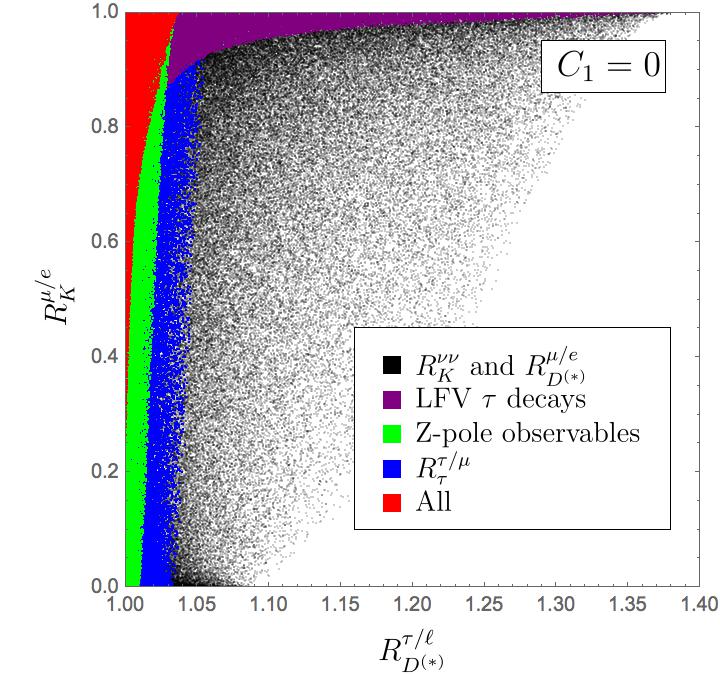}
\end{minipage}~~~~~~~~~~~~~~~
\begin{minipage}{0.45\textwidth}
\includegraphics[width=\textwidth]{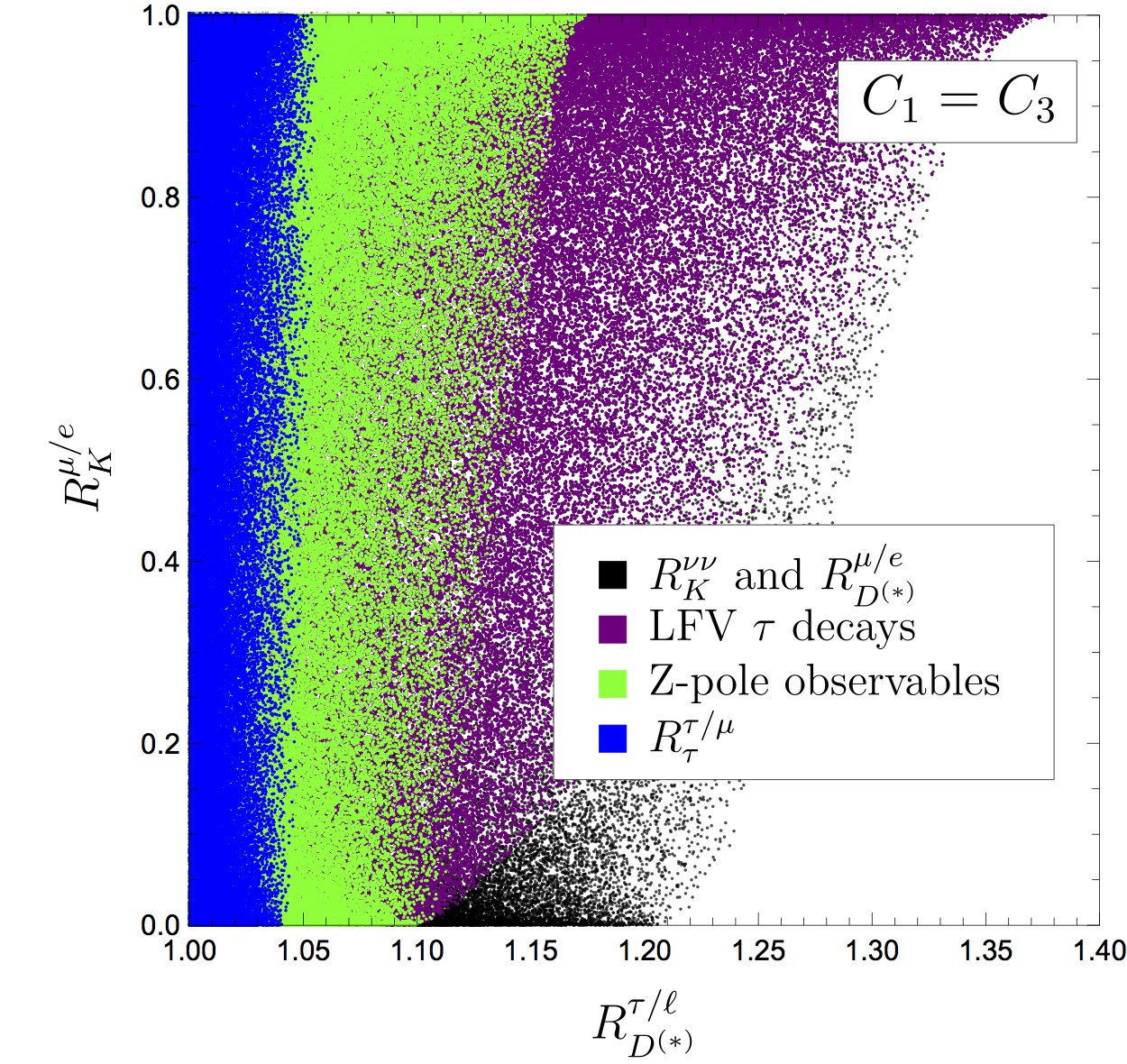}
\end{minipage}
\caption{Impact of constraints arising from quantum corrections in the benchmark model, for two different $C_{1}$ vs.~$C_3$ configurations (\emph{left}: $C_1=0$, \emph{right}: $C_1 = C_3$).
For $C_1=C_3$, simultaneously imposing all bounds is actually equivalent to impose $R_\tau^{\tau/\mu}$ alone.
In the scan the parameters varied in the following ranges: $C_{1,3}/\Lambda^2  \in \{-4,4\}~{\rm TeV}^{-2}$, $\Lambda \in \{1,10\} {\rm TeV}$, $ |\lambda^{d,e}_{23} | \in \{0,0.5\}$. All bounds refer to $2\sigma$ uncertainties. Figures from ref. \cite{Feruglio:2016gvd,Feruglio:2017rjo}.
				\label{fig_numerical1}}
\end{figure}

\begin{figure}[h!]
\centering
\begin{minipage}[t]{0.48\textwidth}
\includegraphics[width=\textwidth]{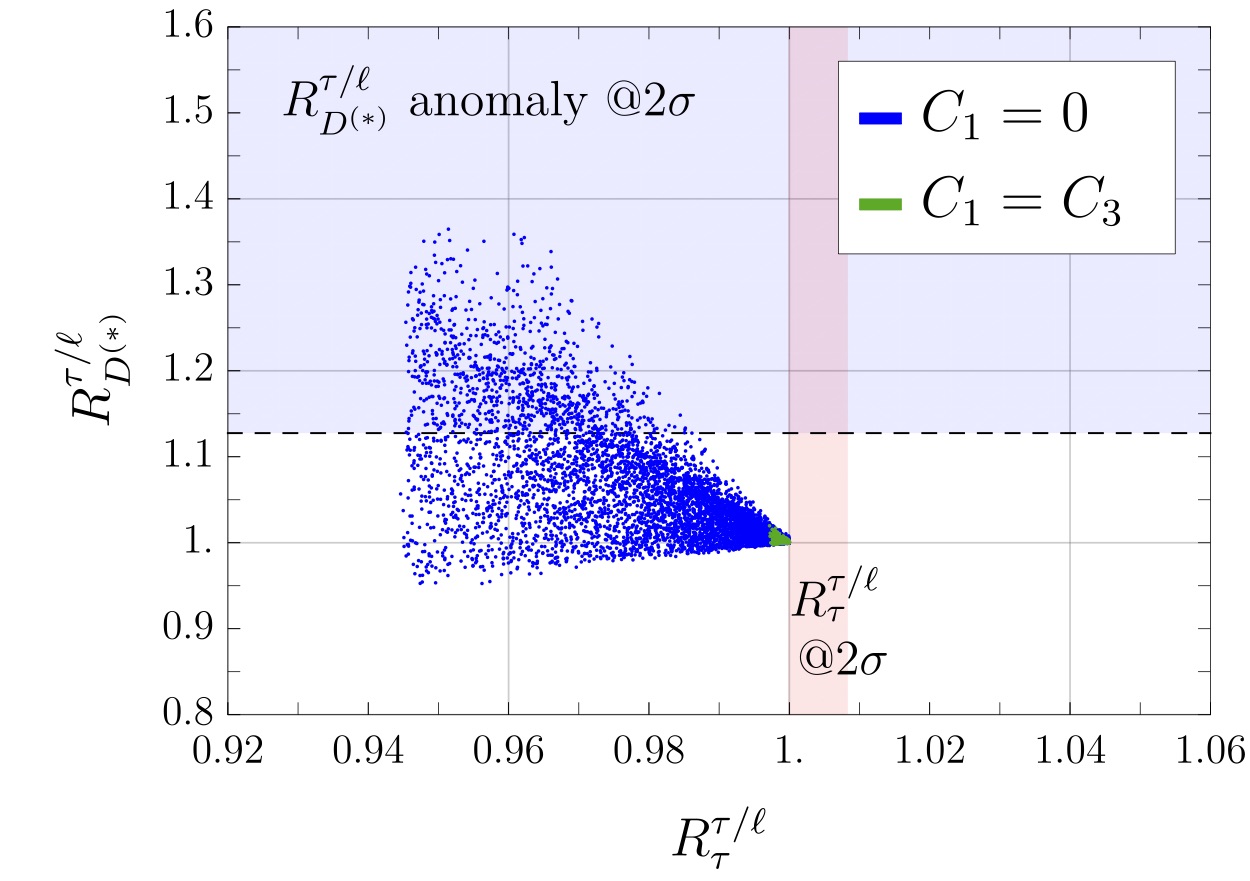}
\end{minipage}
~~~
\begin{minipage}[t]{0.46\textwidth}
\includegraphics[width=\textwidth]{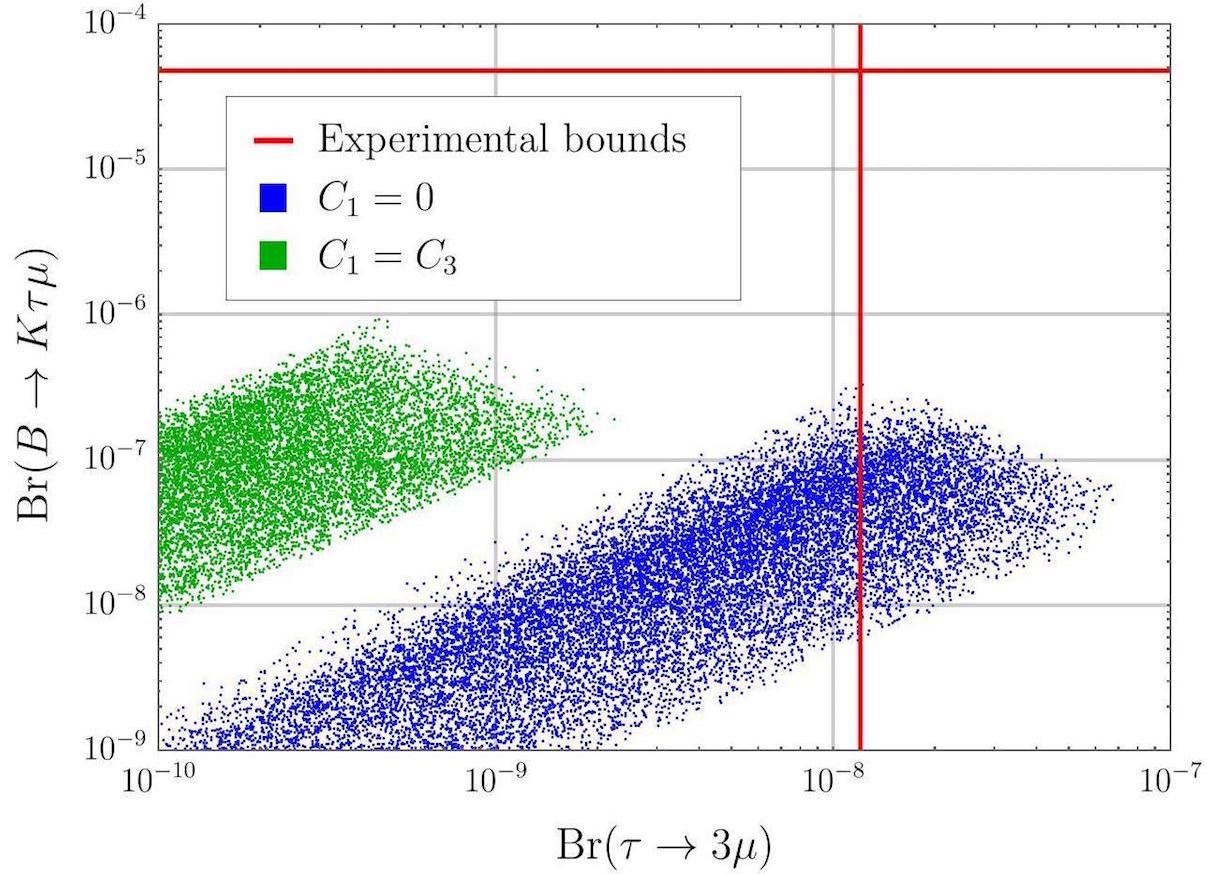}
\end{minipage}

\caption{{\it Left }: 
Correlation between $R_\tau^{\tau/\ell}$ and $R_{D^{(*)}}^{\tau / \ell}$ predictions when scanning the parameter space of the model. In the scan the parameters varied in the following ranges: $C_{1,3}/\Lambda^2  \in \{-4,4\}~{\rm TeV}^{-2}$, $\Lambda \in \{1,10\} {\rm TeV}$, $ |\lambda^{e}_{23} | \in \{0,0.5\}$, $ \lambda^{d}_{23}  \in \{-0.2, -0.01\}$. The $2\sigma$ lower limit for the $R_{D^{(*)}}^{\tau / \ell}$ anomaly and the
	     combined $2\sigma$ bounds of $R_\tau^{\tau/\mu}$ and $R_\tau^{\tau/e}$ are also shown.
{\it Right }: Correlation ${\rm Br}(\tau \to 3 \mu)$ vs.~${\rm Br}(B \to K \tau \mu)$ (${\rm Br}(\tau \to 3 \mu)$ vs.~${\rm Br}(\tau \to \mu \rho)$)  within our model, while satisfying all other bounds but $R_{D^{(*)}}^{\tau / \ell}$, for two different $C_{1}$ vs.~$C_3$ configurations. All bounds refer to $2\sigma$ uncertainties. Figures from ref. \cite{Feruglio:2016gvd,Feruglio:2017rjo}.
\label{fig_numerical2}}
\end{figure}

Crossing the electroweak scale, the gauge bosons $W/Z$, the Higgs boson and the top quark are integrated out
and the running of the effective Lagrangian continues to lower scales. When $\mu=m_\tau$ such a Lagrangian contains
a set of purely leptonic dimension-6 operators violating lepton flavour and LFU. Such operators contribute to leptonic
decays of the $\tau$. As far as tests of LFU are concerned, see eq. (\ref{Rtau}), by keeping only linear terms in the NP contributions we have
\be
R^{\tau/\ell}_\tau  
\approx 1 + \frac{0.008 \, C_3}{\Lambda^2({\rm TeV^2})}
\label{eq:tau_LFU}
\,,
\ee
to be compared with the present result, $R^{\tau/e}_\tau = 1.0060 \pm 0.0030$ and $R^{\tau/\mu}_\tau = 1.0022 \pm 0.0030$~\cite{Pich:2013lsa}.
Violation of LFU in $\tau$ decays is closely related to LFV, which can manifest in several channels \cite{Lusiani}. For example
this framework predicts $\tau\to \mu\ell\ell$ at rates close to the present experimental bounds.
If $(C_1-C_3) \approx \mathcal{O}(1)$, the leading effects on $\mathcal{B}(\tau\to \mu\ell\ell)$ are proportional to $m_t^2/\Lambda^2$ 
and the following numerical estimate applies:
\begin{equation}
\mathcal{B}(\tau\to 3\mu) \approx 5\times 10^{-8} \,
\frac{(C_1-C_3)^{\,2} }{\Lambda^4({\rm TeV^{4}})} 
\left(\frac{\lambda^e_{23}}{0.3}\right)^2 \,,
\label{eq:tau3mu_num}
\end{equation}
to be compared with the current experimental bound $\mathcal{B}(\tau\to 3\mu) < 1.2 \times 10^{-8}$~\cite{Amhis:2014hma}.
Similarly, the decays $\tau\to\mu\rho$ and $\tau\to\mu\pi$ are predicted with branching ratios:

\bea
\mathcal{B}(\tau\to\mu\rho) &\approx& 5 \times 10^{-8} \frac{(C_1 - 1.3 \,C_3)^2}{\Lambda^4({\rm TeV^{4}})} 
\left(\frac{\lambda^e_{23}}{0.3}\right)^2\,,\\
\mathcal{B}(\tau\to\mu\pi) 
&\approx& 8 \times 10^{-8} \, \frac{(C_1-C_3)^2}{\Lambda^4({\rm TeV^{4}})} 
\left(\frac{\lambda^e_{23}}{0.3}\right)^2
\,,
\label{eq:taumurho_num}
\eea
where the current bounds are $\mathcal{B}(\tau\to \!\mu\rho) < 1.5 \times 10^{-8}$ and $\mathcal{B}(\tau\to\mu\pi) < 2.7\times 10^{-8}$~\cite{Amhis:2014hma}.

The overall impact of these constraints is displayed in fig. \ref{fig_numerical1}, for two typical choices of the parameters 
$C_{1,3}$, namely $C_1=0$ and $C_1=C_3$.
The black dots are allowed by tree-level semileptonic bounds, i.e. those discussed in section 2.
When $C_1=0$, LFU violation in $Z$ decays provides the most powerful constraint, while for $C_1=C_3$ the constraint coming from 
$R^{\tau/\mu}_\tau$ is the strongest one. In both cases values of $R_{D^{(*)}}^{\tau / \ell}$ exceeding $1.05$ are strongly disfavoured.
In fig.~\ref{fig_numerical2}, left panel, we reach the same conclusion by the comparing the $R_{D^{(*)}}^{\tau / \ell}$ prediction 
with the most challenging LFUV observable, $R^{\tau/\ell}_\tau$.
Finally, the right plot of fig.~\ref{fig_numerical2} shows the LFV predictions of the benchmark model.
In this plot all bounds, but the $R_{D^{(*)}}^{\tau / \ell}$ anomaly, are satisfied. 
We see that the process $\tau \to 3 \mu$ is preferred over $B \to K \tau \mu$ to prove LFV effects in this scenario,
due both to the closeness of the predicted ${\rm Br}(\tau \to 3 \mu)$ to the present experimental bound and to the expected improvements 
of such bound in the near future. 
In conclusion, a simultaneous explanation of both the $R_{D^{(*)}}^{\tau / \ell}$ and  $\RKss$ anomalies is strongly disfavoured 
in the benchmark scenario, where NP at the TeV scale affects left-handed currents and the third fermion generation.
\section{Ways Out}
The above conclusions are essentially unchanged moving in a more general setup defined by the 
most general set of (current$\times$current) gauge-invariant semileptonic operators 
involving only the 3rd generation \cite{Cornella:2018tfd}:
\bea
\label{Lgen}
{\cal L}_{NP}^0(\Lambda)&=&\frac{1}{\Lambda^2}
\left(C_1~ \bar q'_{3L}\gamma^\mu q'_{3L}~ \bar \ell'_{3L}\gamma_\mu \ell'_{3L}
+C_3~\bar q'_{3L}\gamma^\mu \tau^a q'_{3L}~ \bar \ell'_{3L}\gamma_\mu \tau^a \ell'_{3L}\right.\nn\\
&+&\left.C_4~ \bar d'_{3R}\gamma^\mu d'_{3R}~ \bar \ell'_{3L}\gamma_\mu \ell'_{3L}
+C_5~ \bar d'_{3R}\gamma^\mu d'_{3R}~ \bar e'_{3R}\gamma_\mu e'_{3R}
+C_6~ \bar q'_{3L}\gamma^\mu q'_{3L}~ \bar e'_{3R}\gamma_\mu e_{3R}
\right)
\eea
Also in this case  we find that the most relevant effects of quantum corrections are the modification of the leptonic $W/Z$ couplings and the generation of a purely leptonic effective Lagrangian, both involving LFU violation and LFV at the same time.
For example, a combination of Wilson coefficients favoured by global fits to NC semileptonic $B$-decays 
is realized by choosing $C_1+C_3=C_6$ and $C_4=C_5=0$. This choice reproduces at low energies 
a NC operator product of a $V-A$ quark current and a V charged-lepton current.
A numerical analysis of this particular example and a more general scan over the full parameter space of the model based on eq. (\ref{Lgen}) confirm and reinforce the conclusion that the stringent experimental bounds on $Z$-pole observables 
and $\tau$ decays forbid a simultaneous explanation of NC and CC anomalous data,
at least within the reasonable, though restrictive, assumptions of the benchmark scenario.

There are more general conditions under which this negative result can be evaded.
A first possibility is that the leading logarithmic contributions arising from the RGE analysis are partially/fully cancelled by
finite terms arising at the scale $\Lambda$ in a UV complete model. These finite terms can be described by a set of purely leptonic operators,
or operators involving leptons and $W/Z$ bosons, which have not been included in the framework discussed above.
Though logically possible, we think that this circumstance is rather unlikely, since it would require a tuning of two completely independent 
sets of contributions in a variety of physical observables.

A second possibility arises by allowing for a more general flavour pattern of the NP effects. A crucial property of the benchmark scenario
is that the couplings to the $c$ quark and to the $\mu$ lepton arise entirely from a mixing with the third generation.
Indeed, by allowing from the beginning NP operators affecting directly the second generation,
$B$-anomalies can be explained making use of a larger scale, $\Lambda>1$ TeV \cite{Buttazzo:2017ixm}.
On the one hand, a positive $\lambda^d_{23}$ of order $0.1$ allows to raise the NP scale $\Lambda$ and decouple the radiative effects to a negligible level. The solution to the neutral current anomalies now requires $\lambda^e_{22}<0$, which is incompatible with the flavour pattern assumed in our benchmark,
but possible in a more general context. On the other hand, the new value of $|\lambda^d_{23}|$, much larger than the one considered above, can generate a tension in the phenomenology of $|\Delta F|=2$ transitions.

The correct choice of the NP flavour pattern is an important ingredient in model building. Recently
several UV-complete models have been constructed \cite{Assad:2017iib,DiLuzio:2017vat,Bordone:2017bld,Barbieri:2017tuq,Blanke:2018sro,Greljo:2018tuh,Bordone:2018nbg}, containing in their spectrum a color-triplet, weak-singlet, vector LQ: $U_1=(3,1,+2/3)$ . This state does not contribute to proton decay and naturally arises in a Pati-Salam quark-lepton unification. If $U_1$ couples to left-handed fermions, its exchange gives rise to semileptonic operators of the type $(\bar q'_{pL}\gamma^\mu \tau^a q'_{rL})~ (\bar \ell'_{sL}\gamma_\mu \tau^a \ell'_{tL})$ and $(\bar q'_{pL}\gamma^\mu q'_{rL})~ (\bar \ell'_{sL}\gamma_\mu  \ell'_{tL})$, with equal Wilson coefficients. The main problem of the simplest realization of this scenario is
the large contribution to flavour changing transitions, which pushes the lower limit on the $U_1$ mass to 100 TeV.
Suitable non-minimal modifications of the model are needed in order to lower the $U_1$ mass to the required level.

Finally, even in the benchmark scenario considered at the beginning, both the NC and the CC $B$-anomalies can be
individually accommodated by appropriate choices of the parameters. The solutions exploit the fact that RGE effects
are inversely proportional to $\Lambda^2$ and become negligible as soon as $\Lambda$ is greater than few TeV.
For instance, CC anomalies are explained by $\theta_d\approx 1$, $\theta_e\ll \alpha_{em}$ and $\Lambda\approx 5$ TeV,
while NC anomalous data can be reproduced by $\theta_d\approx 1$, $\theta_e\approx 1$ and $\Lambda\approx 30$ TeV.
All these considerations show that the negative conclusion drawn within the benchmark scenario does not
constitute a no-go theorem, but rather suggests what are the required ingredients for a successful model.
\section{Relation to $(g-2)_\mu$}
It would be interesting to establish a relation between the hints of LFU violation in $B$-decays and the discrepancy in $(g-2)_\mu$.
The extra contribution needed to fix $(g-2)_\mu$ should have a size similar to the electroweak contribution, which
scales as $m_\mu^2/m_W^2$. If the source of this extra contribution is the exchange of an heavy particle of mass $M\approx 1$ TeV,
some enhancement is needed. This enhancement can be provided by the special features of scalar LQ,
such as $S_1=(\bar{3},1,+1/3)$ and $R_2 =(3,2,+7/6)$. Such states can couple to both left-handed and right-handed quarks
and this results into chirally-enhanced contributions to dipole transitions. If the quark exchanged in the loop is the top quark,
we have $(g-2)_\ell\propto m_\ell m_t/M^2$ and $\Gamma(\ell\to\ell'\gamma)\propto \alpha_{ew} m_\ell^3 m_t^2/M^4$ \cite{Bauer:2015knc,ColuccioLeskow:2016dox}.
A contribution to $(g-2)_\mu$ of the right size can arise even for $M\approx 1$ TeV and coupling constants in the weak coupling regime. Many models addressing $B$-anomalies include $S_1$ and/or $R_2$ in their spectrum. If the LQ couples mainly to the top quark and to the second lepton generation, there is a direct relation between $\delta(g-2)_\mu=+3\times 10^{-9}$ and a deviation $\delta\mathcal{B}(Z\to \mu^+\mu^-)\approx 10^{-4}$ in the branching ratio of $Z$ into $\mu^+\mu^-$. More interestingly, if the LQ and the top quark have similar couplings to leptons of the second and third generation, than the framework predicts rates for $\tau\to\mu\gamma$ close to the present experimental sensitivity. A model-independent relation between
the CC $B$-anomaly and $\delta(g-2)_\mu$ can be found in the framework of an effective field theory dominated by scalar and tensor
dimension-6 operators \cite{Feruglio:2018fxo}.

\section*{Ackowledgements}
I am very grateful to the organizers of the Beauty 2018 Conference for inviting me to this very interesting and stimulating event.
I warmly thank Claudia Cornella, Andrea Pattori and Olcyr Sumensari for the pleasant collaboration
on the topics of this talk. A special thank goes to Paride Paradisi, who introduced me to $B$-phsyics with his enthusiasm and passion. This work and my participation to the Conference were supported in part by the MIUR-PRIN project 2010YJ2NYW and by the European Union network FP10 ITN ELUSIVES and INVISIBLES-PLUS (H2020- MSCA- ITN- 2015-674896 and H2020- MSCA- RISE- 2015- 690575).


\end{document}